# Tera-sample-per-second Real-time Waveform Digitizer

Yan Han[*], Ozdal Boyraz, and Bahram Jalali

Department of Electrical Engineering, University of California, Los Angeles, CA 90095.

E-mails: yanhan@creol.ucf.edu, Jalali@ucla.edu

## ABSTRACT

We demonstrate a real-time transient waveform digitizer with a record 1 TSa/s (Tera-Sample/sec) sampling rate. This is accomplished by using a photonic time stretch preprocessor which slows down the electrical waveform before it is captured by an electronic digitizer.

---

[*] Present address: College of Optics and Photonics: CREOL & FPCE, University of Central Florida, 4000 Central Florida Blvd., Orlando, FL 32816

# Tera-sample-per-second Real-time Waveform Digitizer

Digital processing and storage of information is a ubiquitous function encountered in virtually every science and engineering discipline. Since signals originating in the physical world are analog, Analog-to-Digital Converters (ADC), otherwise known as digitizers, play a central role. The most challenging signals to digitize are ultrafast transients that are non-repetitive. These waveforms arise in a number of applications including (i) diagnostic tool in particle accelerators that probe fundamental building blocks of nature; (ii) X-ray free electron lasers[1]; (iii) study of EMP (electromagnetic pulse) weapons[2]. Digitizers with real-time capability are required because the non-repetitive nature of these signals renders traditional sampling oscilloscopes useless.

The sampling rate of conventional electronic ADCs, such as ones used in high speed instruments, is limited by the speed of the electronic circuitry, and in the case where the interleaving architecture is used, by the mismatches in the multi-channel ADC array[3,4]. While the performance of electronic ADCs continues to improve, the sampling rate of a state-of-the-art system is currently about 20 GSa/s with ~5 ENOB (Effective Number Of Bits). Achieving TSa/s performance is clearly beyond the reach of conventional approaches. One potential solution to overcome the electronic bottleneck is to use photonic pre-processing. In particular, the photonic time stretch approach has proven to be an effective way to extend the sampling rate and the bandwidth. Here, the high speed transient waveform is first slowed down and then captured by a conventional electronic digitizer[5,6]. In this paper, we demonstrate such a system that achieves a sampling rate of 1 TSa/s. This system consists of a 50x time stretch pre-processor and a 20 GSa/s electronic digitizer. To the best of our knowledge, this is the first time that the real-time digitization at 1 TSa/s has been achieved.

The time stretch preprocessing, shown in Fig. 1, consists of three steps: time-to-wavelength transformation, wavelength domain processing, and wavelength-to-time transformation. Time-to-wavelength transformation occurs when the electrical signal modulates the intensity of a linearly-chirped optical pulse. At the output of modulator, the input signal's time scale is linearly mapped onto the optical wavelength. The second and third steps occur simultaneously when the waveform is broadened as it travels through the second dispersive optical medium and is subsequently photodetected.

The major obstacle to achieving a stretch factor of 50x is to overcome the frequency fading that is associated with dispersive propagation. This phenomenon has been described in details elsewhere[5,6]. Briefly, it occurs due to the dispersion-induced interference between the two modulation-sidebands. We overcome this problem using the recently proposed phase diversity technique, where two stretched waveforms with complementary fading characteristics are realized and combined to eliminate the bandwidth limitation[7]. Another practical concern is the large loss of dispersive fiber required to achieve such a large stretch factor. This problem is mitigated by the judicious use of optical amplification in such a manner as to optimize the overall signal to noise ratio while avoiding degradation from optical nonlinearity.

The block diagram of experimental setup for a 50x photonic time stretch preprocessor is shown in Fig. 2. Ultrashort and broadband optical pulses are produced, with repetition period $T$=200ns, through the continuum generation process. Around 15 nm of the optical spectrum at (centered at 1590 nm) is sliced using an optical bandpass filter and is used in the experiment. After being linearly chirped by propagation in a spool of dispersive fiber, a given pulse enters the electro-optic modulator where it captures the electrical transient. The chirped optical pulse has finite amplitude variations resulting in distortion of the captured signal. The distortion is removed by using a reference waveform and digital filtering[6]. The modulator has two outputs with complimentary fading characteristics, and after delaying one by $T/2$, the outputs are interleaved and stretched together in the second dispersive fiber. The first fiber has a total dispersion of $D_1$ = -101 ps/nm, creating a chirped optical pulse with around 1.5 ns duration. The second fiber has a total dispersion of $D_2$ = -4921 ps/nm. The stretch factor is give by $(D_1+D_2)/D_1 = 50$[5,6]. To minimize the noise contributed by the optical amplifiers, the waveform is filtered by an optical bandpass filter before photodetection. The detected waveform is subsequently captured by a Tektronix TDS7404 (4GHz analog bandwidth, 20GSa/s) real-time digitizer. The effective sampling rate of this system is 20 GSa/s x 50 = 1 TSa/s and the intrinsic input analog bandwidth is 4 GHz x 50 = 200 GHz, although, in the experiments reported here, the electro-optic modulator places a limit of around 50 GHz.

Figure 3 shows the real-time capture of a 48 GHz tone at 1 TSa/s. The measured voltage Full Width Half Maximum (FWHM) time aperture is 1.1 ns. The time aperture determines the length of transients that can be captured. The measured stretch factor is 50.5. To obtain the average SNR, 200 such real-time measurements are preformed. Over a 10 GHz digitally filtered bandwidth centered at 48 GHz, the mean and standard deviation of SNR in each of 200 measurements is $22.7 \pm 1.08$ dB corresponding to $3.5 \pm 0.2$ ENOB. Measurements at other input frequencies with the same bandwidth resulted in up to 4.2 ENOB.

The digitizer performance is also measured in the frequency domain. Fig. 4 shows the digital spectrum of a measured 26 GHz signal and a 42 GHz signal, respectively. The Hanning window was used before discrete Fourier transform. We note that no digital filter was used to obtain the data in Fig. 4, hence the input analog bandwidth of these spectrums is 0 – 200 GHz. The $2^{nd}$ harmonic distortion tone is observed in Fig. 4. These arise due to nonlinear electrooptic modulation that is affected by memory effects in the fiber, a phenomenon that has been previously predicted by theory[6]. Post-nonlinear-compensation performed in the digital domain similar to broadband power amplifier linearization in wireless communications can potentially mitigate this type of distortion. The highest spurious peak appearing at 125 GHz is due to the electronic digitizer and is independent of the time stretch processor.

Although the TSa/s digitizer has an intrinsic bandwidth of 200 GHz, the experimental input bandwidth is limited by the bandwidth of the electro-optic modulator to approximately 50 GHz. One method to effectively increase the modulation bandwidth is to use harmonic modulation. Conventionally, the MZ modulator is biased at the quadrature point, at which even harmonics are suppressed. If a MZ modulator is biased at the peak or null of transmission, fundamental frequency and odd harmonics, instead, are suppressed. The lowest frequency component is the second harmonic. Hence, an 80 GHz sinusoid modulated optical pulse can be generated from a 40 GHz input. The digital magnitude spectrum of captured 80 GHz signal is shown in Fig. 5. The clean spectral line at 80 GHz demonstrates the potential of time stretch processor beyond 50 GHz if an MZ modulator with a larger bandwidth is used in the setup. The spurious tones at 40 and 120 GHz are from the under-suppressed fundamental frequency and the 3$^{rd}$ harmonic.

Optical nonlinearity may distort the stretched signal. It is especially important in TSa/s experiment, in which multiple optical amplifications are needed to compensate fiber loss. Improper amplifier design may significantly distort signal. An example of distortion pattern is shown in Fig. 6. The distortion (amplitude reduction) is most severe at the center of pulse, where the pulse has highest peak power. This measured distortion pattern is verified by the numerical simulation and is the result of interaction between fiber dispersion and optical nonlinearity. When the optical pulse is modulated by a sinusoid signal, optical nonlinearity introduces an intensity dependent phase shift in propagation. Dispersion causes phase-to-amplitude conversion and hence results in distortion of signal amplitude[6]. Because optical power is highest at the center of time window, the extent of nonlinear distortion varies across the time window and introduces the distortion pattern observed in Fig. 6.

In summary, the photonic time stretch technique has been used to realize a real-time transient digitizer with the record one TSa/s sampling rate. The system has a FWHM time aperture of 1.1 ns. Measured over a 10 GHz bandwidth, the ENOB of the system ranges from 3.5-4.2 bits depending on the input signal frequency.

Acknowledgement: This work was supported by DARPA.

**Fig. 1. Block diagram of the photonic time stretch preprocessor.**

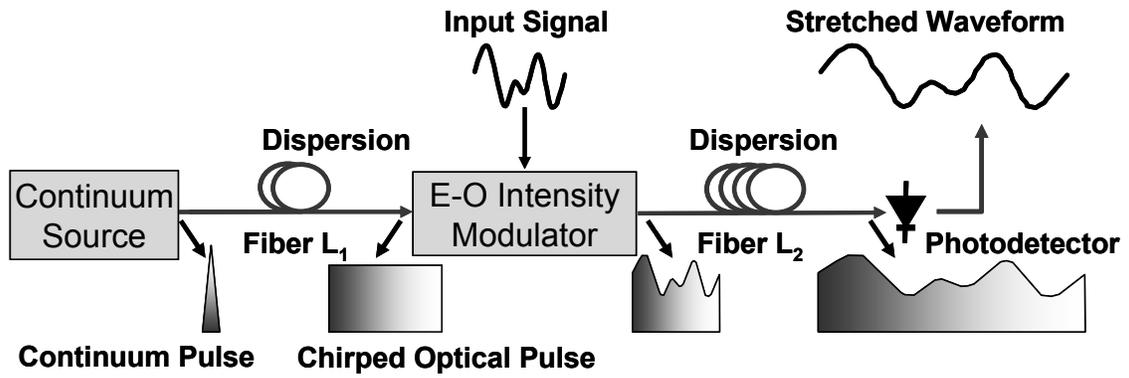

Fig. 2. Experiment setup of TSa/s digitizer with a 50x photonic time stretch preprocessor. EDFA: Erbium-Doped Fiber Amplifier.

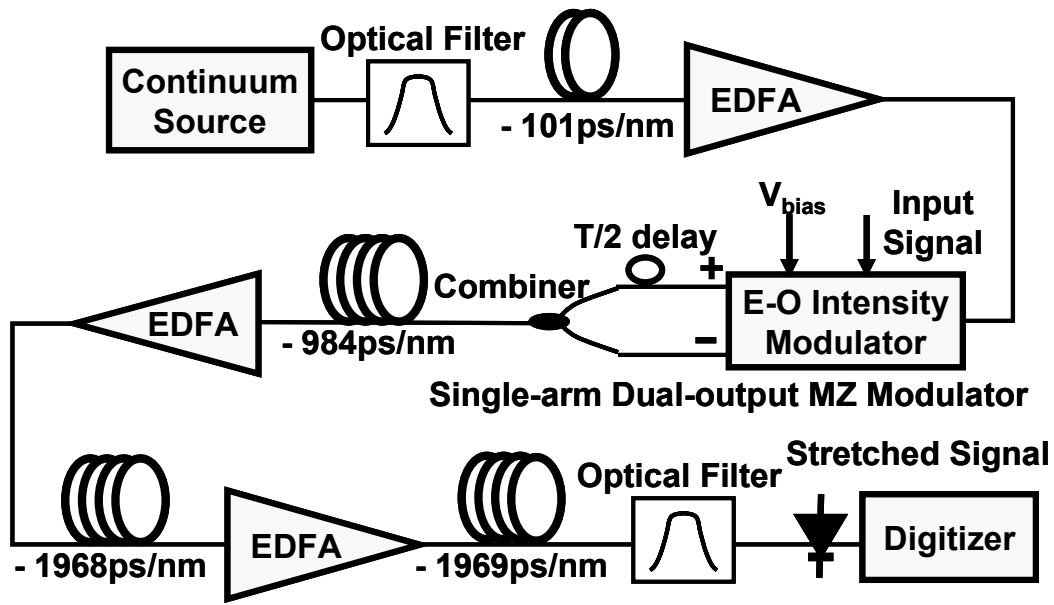

Fig. 3. Digitization of a 48 GHz sinusoid at 1 TSa/s. Lines are obtained using standard sine curve fitting.

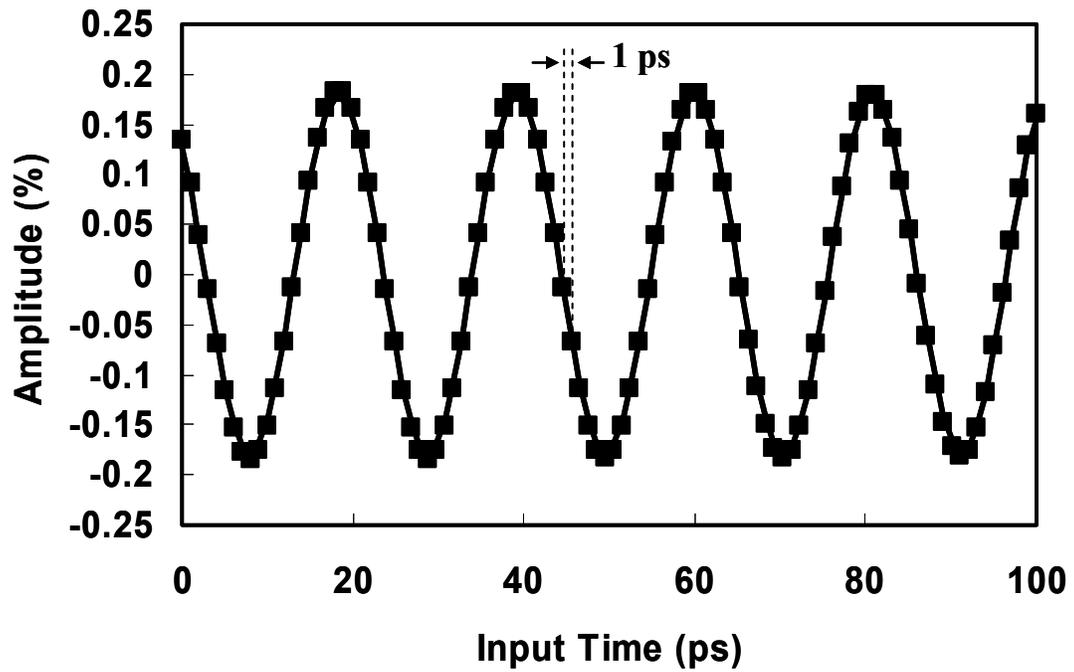

**Fig. 4.** The digital magnitude spectrum of captured signals at 26 GHz (solid line) and 48 GHz (dashed line).

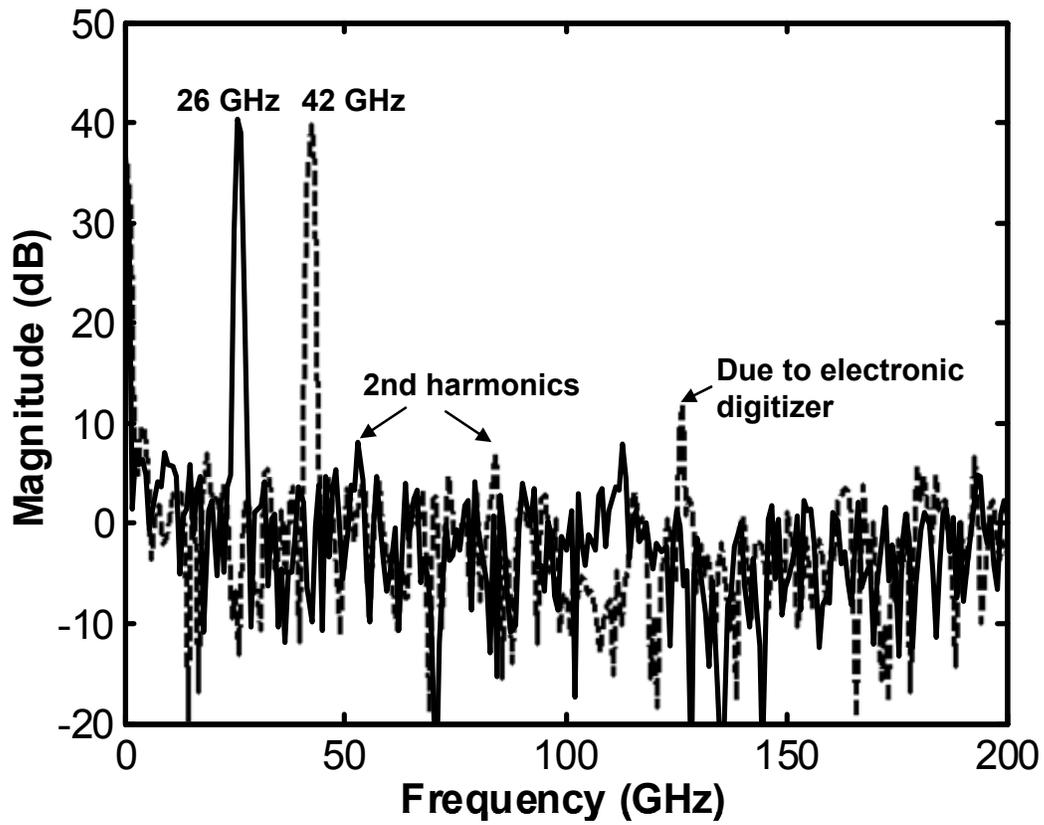

**Fig. 5. The digital magnitude spectrum of captured signals at 80 GHz.**

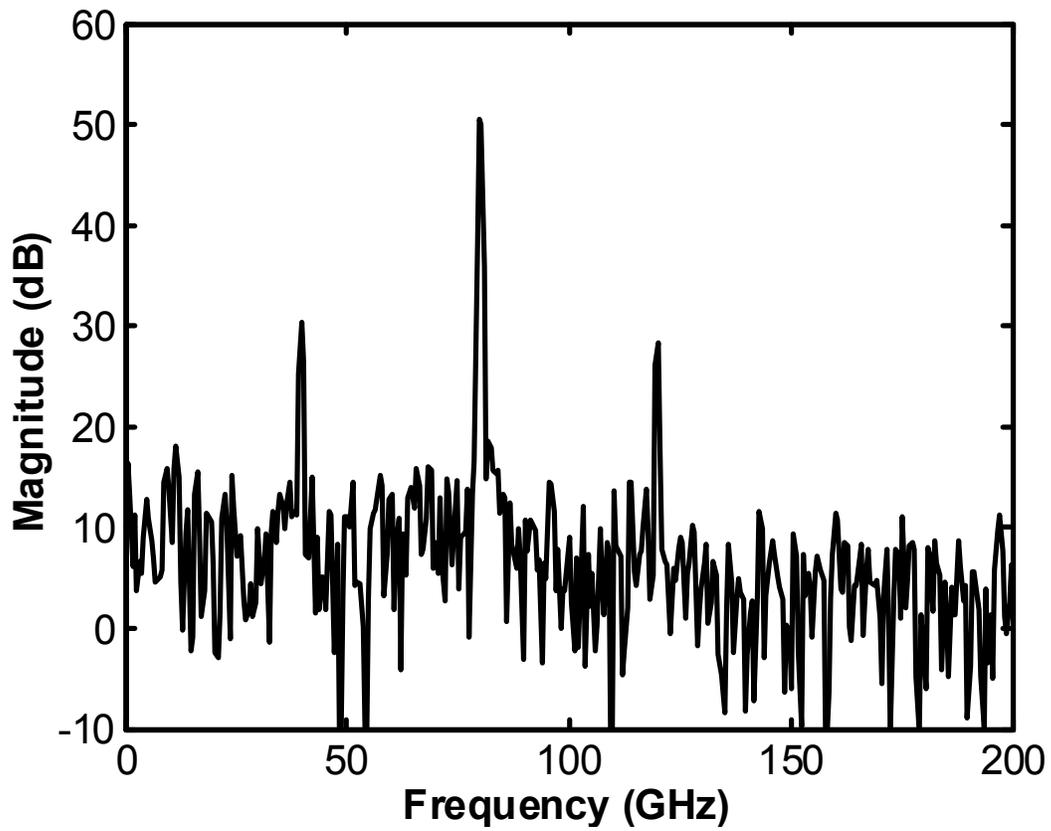

**Fig. 6. Distortion caused by nonlinear optical effect in the dispersive fiber.**

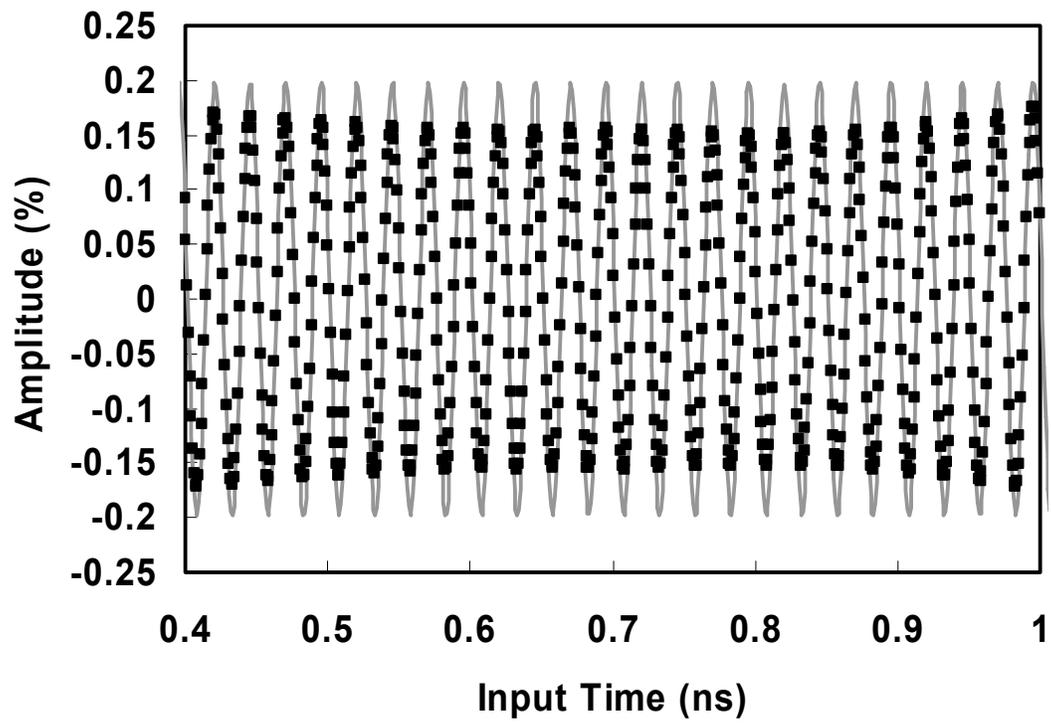